\documentclass[pra,nofootinbib,floats,aps,twocolumn,tightenlines,superscriptaddress]{revtex4-1}

\usepackage{graphicx}
\usepackage{bbold}
\usepackage{mathtools}
\usepackage{hyperref}
\usepackage{commath}
\usepackage{bm}
\usepackage{amsfonts,amsmath,amssymb,amsthm}
\usepackage[usenames,dvipsnames]{xcolor}
\usepackage{subfigure}

\DeclareMathOperator{\Ci}{Ci}

\newcommand{\lam}{\ensuremath{\sqrt{|\Lambda|}}}
\newcommand{\lap}{\ensuremath{\nabla^2}}
\renewcommand*\d[2][]{%
	\mathrm{d}%
	\ifx\relax#1\relax\else
	\rule{-0.02em}{1.5ex}^{#1}\rule{0.08em}{0ex}\!
	\fi
	#2\,
} 
\newcommand{\ket}[1]{| {#1} \rangle}
\newcommand{\bra}[1]{\langle {#1} |}
\newcommand{\tr}{\text{Tr}}
\renewcommand{\Re}[1]{\operatorname{Re}{#1}}

\renewcommand{\Ci}{\text{Ci}}
\newcommand{\ii}{\mathrm{i}}

\begin{abstract}
	We analyze the exchange of information in different cosmological backgrounds when sender and receiver are timelike separated and communicate through massless fields (without the exchange of light-signals). Remarkably, we show that the dominance of a cosmological constant makes the amount of recoverable information imprinted in the field by the sender extremely resilient: it does not decay in time or with the spatial separation of sender and receiver, and it actually increases with the rate of expansion of the Universe. This is in stark contrast with the information carried by conventional light-signals and with previous results on timelike communication through massless fields in matter dominated cosmologies. 
\end{abstract}

\begin{document}

	\title{The information carrying capacity of a cosmological constant}
	
	\author{Petar Simidzija}
	\affiliation{Institute for Quantum Computing, University of Waterloo, Waterloo, Ontario, N2L 3G1, Canada}
	
	\author{Eduardo Mart\'in-Mart\'inez}
	\affiliation{Institute for Quantum Computing, University of Waterloo, Waterloo, Ontario, N2L 3G1, Canada}
	\affiliation{Department of Applied Mathematics, University of Waterloo, Waterloo, Ontario, N2L 3G1, Canada}
	\affiliation{Perimeter Institute for Theoretical Physics, Waterloo, Ontario N2L 2Y5, Canada}

	\maketitle

	\section{Introduction\label{sec:intro}}

	In recent years, there have been a number of results highlighting the relationship between the physics of information and fundamental topics in quantum field theory and gravitation. For example, quantum entanglement in vacuum fluctuations has been linked to phenomena like Hawking radiation and the Unruh effect~\cite{Hotta2015}. Entanglement in the vacuum state of a quantum field can also be viewed as a resource in protocols of \textit{quantum energy teleportation}~\cite{Hotta2009,Hotta2011}, and can be \textit{harvested}~\cite{Valentini1991,Reznik2003,Reznik2005,Franson2008,Lin2009,Pozas2015,Pozas2016}, or \textit{farmed}~\cite{Martin-Martinez2013a}, by particle detectors which locally couple to the field. These detectors can become entangled with one another, even if they are spacelike separated.  Interestingly, entanglement harvesting has been proven to be sensitive to spacetime curvature~\cite{Steeg2009,Nambu2011,Hu2013,Martin-Martinez2012,Martin-Martinez2014,Tian2014,Salton2015} and even spacetime topology~\cite{Martin-Martinez2016}. However, although two spacelike separated detectors may become correlated just out of their interaction with the vacuum, superluminal broadcasting of information between them is, of course, not possible. 
	
	In this context, it is relevant to ask what is the information carrying capacity of a quantum field. From a fundamental point of view, when we want to transmit information through a quantum field---whether in telecommunication or in an attempt to gather information about the early Universe---a necessary (but not sufficient) condition for communication is that the field commutator between the spacetime events of sending and receiving the message, does not vanish~\cite{Cliche2010,Benincasa2014,Martin-Martinez2015}. 
	
	The (expectation value of the) commutator of a  quantum field is given by the classical radiation Green's function~\cite{Blasco2016} (difference between the retarded and the advanced Green's functions). In this regard, the strong Huygens principle~\cite{McLenaghan1974} states that the support of the radiation Green's function of a massless field is restricted to lightlike separated events, implying that only lightlike signaling is possible. This is consistent with our everyday intuition: if we beam an empty chair with a laser and no one is there to receive the message, the information is gone, and it is not recoverable by a late receiver that sits on the chair the next day. However, for general spacetimes, the strong Huygens principle can be violated~\cite{McLenaghan1974,Blanchet1988,Blanchet1992,Faraoni1992,Bombelli1994,Gundlach1994,Faraoni1999}. In these cases, massless field commutators can have support for timelike separated events. In fact, even for a simple massless scalar field, these violations are extremely common: the strong Huygens principle is violated in almost any curved spacetime, and in flat spacetimes of (1+1) and (2$n$+1)-dimensions \cite{McLenaghan1974,Jonsson2015,Blasco2016}.
	
	Note that violations of the strong Huygens principle are not enough to guarantee timelike separated observers the ability to communicate. It was shown in~\cite{Jonsson2015} however, that if, additionally, the observers operate quantum antennas initialized to coherent superpositions of ground and excited eigenstates, a timelike signaling protocol can be established. Furthermore, this protocol allows for the possibility of broadcasting a message to an arbitrary number of timelike receivers, with the energy cost of transmitting the message being paid for by the receivers themselves. Because of this, this protocol received the name \textit{quantum collect calling}.
	
	This method of information broadcasting has been studied in great detail  in~\cite{Blasco2015,Blasco2016} for a polynomially expanding, Fridmann-Robertson-Walker (FRW) cosmology generated by matter, in which reside two comoving observers: an early time signal emitter, Alice, and a later time receiver, Bob.  It was shown that in this universe the timelike communication channel capacity is independent of the observers' spatial separation, but decays as the instant that Bob attempts to retrieve Alice's message (by coupling his antenna to the field) goes further into the future.
	
	In the present work, we analyze the ability of timelike separated observers to communicate in an FRW universe dominated by a cosmological constant, which expands exponentially in comoving time, and we compare this to the matter-dominated case. We consider \mbox{(3+1)-dimensions} and minimal coupling of the massless scalar quantum field to the geometry. We supply Alice with a particle detector with which she can couple to the field, thereby leaving behind information which Bob can recover at a later time by coupling his own detector to the field. 
	
	We will show that timelike communication in an exponentially expanding universe displays unexpected and remarkable features that fundamentally impact the channel capacity. Namely, while we may expect that in an exponential expansion less information will reach Bob than in a polynomial one---due to the information being dispersed more in the faster expanding case---we show that, in fact, the opposite occurs. In the exponentially expanding universe, Bob's ability to recover Alice's message remains the same regardless of how long he waits before reading it out, in stark contrast with the decay present in the polynomially expanding cosmology. What is more, we find that Alice can broadcast more information to Bob the faster the exponential expansion of space is.
	
	Not only does this imply that in principle more information is available to Bob through the timelike channel than through a light signal (which was proven in~\cite{Blasco2015,Blasco2016} to decay with the distance from the source), but it also means that Bob's ability to access Alice's information remains the same no matter how long Bob waits to switch on his antenna.
	
	The outline of this article is as follows: Sec.~\ref{sec:geometry_field} introduces the field-detector setup, along with the background spacetime geometry. In Sec.~\ref{sec:communication}, the communication protocol is defined, and the ability of Alice to signal Bob is quantified through their channel capacity. Sec.~\ref{sec:timelike_communication} particularizes the channel capacity in each cosmology to the case of timelike separated observers, and compares the two models within this causal regime. In this section we also look at the dependence of the channel capacity on whether we keep constant the proper or the comoving distance separating Alice and Bob. In Sec.~\ref{sec:conclusions}, we present our conclusions. Natural units $\hbar=c=1$ are used throughout.
	
	\section{Background setup\label{sec:geometry_field}}
	
	We will consider a spatially flat Friedmann-Robertson-Walker cosmology given by the metric
	\begin{align}
	\dif s^2 &= -\dif t^2 + a(t)^2(\dif r^2 + r^2 \dif \Omega^2) \notag \\
	&= a(\eta)^2(-\dif \eta^2 + \dif r^2 + r^2 \dif \Omega^2), \label{eq:metric}
	\end{align}
	where we define the conformal time $\eta$ in terms of the comoving time $t$ (proper to observers comoving with the Hubble flow, also called \textit{cosmological time}) as \mbox{$\dif \eta = \dif t/a(t)$}. The scale factor $a(t)$ quantifies the spatial expansion of the Universe, and its precise form depends on the stress-energy density which generates the spacetime. We will consider a universe generated by a perfect fluid with density $\rho$ and pressure  $p=w\rho$. Specifically, we will focus on the two cases $w=0$ and $w=-1$, which correspond to (dust) matter and cosmological constant-dominated universes respectively. From the Friedmann equations, we obtain for the matter-dominated case that
	\begin{equation}\label{eq:w=0}
	a(t)=(9\kappa_1 t^2)^{1/3},\qquad \eta(t)=\left( \frac{3t}{\kappa_1} \right)^{1/3},
	\end{equation}
	where $t,\eta\in[0,\infty)$. Doing the same for the cosmological constant-dominated case, we get
	\begin{equation}\label{eq:w=-1}
	a(t)=\kappa_2 e^{\lam t},\qquad \eta(t)=-\frac{1}{\lam\kappa_2}e^{-\lam t},
	\end{equation}
	where $t\in(-\infty,\infty)$ and $\eta\in(-\infty,0)$. $\kappa_1$ and $\kappa_2$ are constants of integration. We see that the matter-dominated universe is born out of a Big-Bang singularity and it experiences a polynomial spatial expansion. On the other hand, the cosmological constant-dominated universe does not originate with a Big-Bang (understood as a cancellation of $a(t)$ for a finite value of $t$) and expands exponentially in comoving time $t$.
	
	Let us introduce a test massless scalar quantum field $\phi$. The equation of motion for the field is
	\begin{equation}\label{eq:wave_eqn}
	(\Box-\xi\mathcal{R})\phi=0,
	\end{equation}
	where $\xi$ is the coupling to the Ricci scalar
	\begin{equation}
	\mathcal{R}=\frac{6}{a^3}\frac{\d[2]{a}}{\dif\eta^2},
	\end{equation}
	and where the d'Alambertian operator in the FRW spacetime is given by
	\begin{equation}
	\Box=-\frac{1}{a^4}\frac{\dif}{\dif\eta}\left(a^2\frac{\dif}{\dif\eta}\right)+
	\frac{1}{a^2}\lap.
	\end{equation}
	Same as in~\cite{Blasco2016}, for computational purposes, we will choose the quantization scheme that corresponds to the adiabatic vacuum (see e.g., \cite{Junker2002,birrell_quantum_1982}). This quantization scheme is particularly useful since the adiabatic vacuum corresponds to the field state for which the creation of particles due to the expansion of spacetime is finite and the smallest  possible \cite{birrell_quantum_1982}. Furthermore, as shown in~\cite{Cortez2011,Fernandez-Mendez2012,Cortez2012}. As rigorously discussed in these papers, for conformally flat compact spacetimes there exist natural criteria that select a unique equivalence class of vacua, which includes the adiabatic vacuum. Notice however, that we do not assume that the field is initially prepared in the adiabatic vacuum, and instead allow the field to be in any (non ill-defined)  state. Indeed, for the same reasons as in~\cite{Jonsson2015,Blasco2015,Blasco2016}, the results in this paper will be independent of the initial state of the field, as long as it is well defined.
	
	It was shown in~\cite{Blasco2016} that if the field is minimally coupled to the curvature ($\xi=0$), two timelike separated observers gain the ability to communicate through the massless field without exchanging field quanta, taking advantage of the violations of the Strong Huygens Principle~\cite{Jonsson2015}. In fact this is true not only for minimal coupling, but rather for any coupling to curvature that breaks conformal invariance. For concreteness, in this paper we will focus on the minimally coupled case.
	
	\section{Communication through detectors coupled to the field\label{sec:communication}}
	Let us consider the following communication scenario: An observer in the early universe, Alice, wants to communicate with an observer living at a later cosmological epoch, Bob. We suppose that both Alice and Bob are fundamental (comoving) observers, meaning that they move with the Hubble flow and experience the Universe to be isotropic through its evolution. This assumption seems reasonable considering that all distant galaxies have small peculiar velocities with respect to local fundamental observers. Significantly, Earth bound observers are nearly fundamental as is evident by the observed dominant isotropy of the cosmic microwave background and of galactic densities on large scales. Hence in the above described picture we can think of ourselves as Bob, trying to detect a signal from an early Universe emitter Alice.
	
	Let us assume that Alice operates a \textit{radio emitter} that locally couples to the field, and Bob a \textit{radio receiver} with which he tries to recover the information encoded in the field by Alice. We will model Alice and Bob's antennas as two-level quantum systems (particle detectors) that couple locally to the quantum field through the Unruh-DeWitt interaction Hamiltonian~\cite{DeWitt1980}
	\begin{equation}\label{eq:HI}
	H_{\textsc{i},v}=\lambda_\nu\chi_\nu(t)\mu_\nu(t)\int\d[3]{\bm{x}} a(t)^3 F_\nu[\bm{x}-\bm{x}_\nu(t),t]\phi[\bm{x},\eta(t)],
	\end{equation}
	(where $\text{d}^3\bm x=r^2\text{d}r\,\text{d}\Omega$ and $\bm x$ is a spatial 3-vector), which has been shown to capture the fundamental features of the light-matter interaction when there is no exchange of orbital angular momentum~\cite{Pozas2016,Alhambra2014}. Here $\nu\in\{\text{A,B}\}$ labels Alice and Bob's detectors, and $\mu_\nu(t)$ is the monopole moment of detector $\nu$,
	\begin{equation}
	\mu_\nu(t)=\sigma_\nu^+e^{i\Omega_\nu t}+\sigma_\nu^-e^{-i\Omega_\nu t}.
	\end{equation}
	$\sigma_\nu^+=|e_\nu \rangle \langle g_\nu|$ and $\sigma_\nu^-=|g_\nu \rangle \langle e_\nu|$ are the $\text{SU}(2)$ raising and lowering operators, with $|g_\nu \rangle$ and $|e_\nu \rangle$ the ground and excited states, separated by an energy gap $\Omega_\nu$. The detector-field coupling (for detector $\nu$) is characterized by the coupling strength $\lambda_\nu$ and the switching function $\chi_\nu$, which for simplicity we consider to be the characteristic function
	\begin{equation}\label{eq:switching}
	\chi_\nu(t)=
	\begin{cases} 
	1  & t\in[T_{i\nu},T_{f\nu}]  \\
	0  & t\not\in[T_{i\nu},T_{f\nu}]
	\end{cases}.
	\end{equation}
	$F_\nu[\bm{x}-\bm{x}_\nu(t),t]$ is a smearing function characterizing the geometry of detector $\nu$, centered along its trajectory $\bm{x}_\nu$. We consider comoving detectors, $\bm{x}_\nu=\text{const}$, and for now keep the detector smearing general.
	
	Let each detector start out in the pure state \mbox{$\rho_{0,\nu}=\ket{\psi_{0,\nu}} \bra{\psi_{0,\nu}}$}, where $\ket{\psi_{0,\nu}}=\alpha_\nu\ket{e_\nu}+\beta_\nu\ket{g_\nu}$, and let the field start out in the arbitrary state $\rho_{0,\phi}$. Hence, the initial state of the system is
	\begin{equation}
	\rho_0=\rho_{0,\textsc{a}} \otimes \rho_{0,\textsc{b}} \otimes \rho_{0,\phi}.
	\end{equation}
	Allowing the system to evolve under the full interaction Hamiltonian $H_\textsc{i}(t)=H_{\textsc{i},\textsc{a}}(t)+H_{\textsc{i},\textsc{b}}(t)$ for a time $T$ results in the state $\rho_{_T}=U\rho_0 U^\dagger$, where U is the time evolution operator
	\begin{equation}
	    U=\mathcal{T}\exp\left[-\ii\int_{-\infty}^\infty \!\!\dif t H_\textsc{i}(t)\right],
	\end{equation}
	and $\mathcal{T}$ denotes time-ordering of the exponential. The final state of Bob's detector is obtained by tracing out the degrees of freedom corresponding to the field and the state of Alice:
	\begin{equation}\label{eq:rho_t_b}
	\rho_{_{T,\text{B}}}=\tr_{\phi,\textsc{a}}(\rho_{_T}).
	\end{equation}
	The excitation probability of Bob's detector at time $T$ is given by~\cite{Martin-Martinez2015,Blasco2015,Jonsson2015,Jonsson2014}
	\begin{equation}\label{eq:Pe}
	P_e=\bra{e_\textsc{b}}\rho_{_{T,\text{B}}}\ket{e_\textsc{b}}
	=|\alpha_\textsc{b}|^2+R+S,
	\end{equation}
	where $R$ is the local correction to the excitation probability of Bob (independent of $\lambda_\textsc{a}$), and $S$ is the signaling term (dependent on $\lambda_\textsc{a}$) that captures the influence of Alice's detector on the excitation probability of Bob~\cite{Martin-Martinez2015,Jonsson2015}. We call $S$ the \textit{signaling contribution} to Bob's excitation probability. A power series expansion in the coupling strengths gives
	\begin{equation}
	S=\lambda_\textsc{a}\lambda_\textsc{b}S_2+\mathcal{O}(\lambda_\nu^4),
	\end{equation}
	where the lowest order term, $S_2$, takes the form\footnote{For a step-by-step derivation of~\eqref{eq:S2}, see~\cite{Martin-Martinez2015}, Eq. (5) to (25).}~\cite{Blasco2015,Blasco2016,Jonsson2015} 
	\begin{align}
	&S_2=4\!\int\!\dif v\!\int\!\dif v' \chi_\textsc{a}(t)\chi_\textsc{b}(t')
	F_\textsc{a}(\bm{x}-\bm{x}_\textsc{a},t) 
	F_\textsc{b}(\bm{x'}-\bm{x}_\textsc{b},t') \notag \\
	&\!\times \!	\Re{(\alpha_\textsc{a}^*\beta_\textsc{a}e^{\ii\Omega_\textsc{a}t})}
	\Re{\Big(\alpha_\textsc{b}^*\beta_\textsc{b}e^{\ii\Omega_\textsc{b}t'}\Big\langle [\phi(\bm{x}_\textsc{a},t),\phi(\bm{x}_\textsc{b},t')]\Big\rangle_{\!\!\rho_{_{0,\phi}}}\!\!\Big)}, \label{eq:S2}
	\end{align}
	with $\dif v=a(t)^3\d[3]{\bm{x}}
	\rule{-0.2em}{0ex}\dif t$ being the FRW volume element. Notice that, since the field commutator is a c-number (multiple of the identity), its expectation value is independent of $\rho_{0,\phi}$.
	
	Let us now, for simplicity, particularize the discussion to the limit of point-like detectors, characterized by the smearing function
	\begin{equation}
	F_\nu(\bm{x},t)=\delta(\bm{x}).
	\end{equation}
	Although the use of detectors in this limit along with sudden switching functions is known to cause UV divergences in the excitation probability~\cite{Louko2008}, $S_2$ was proven to be UV-safe~\cite{Blasco2016,Martin-Martinez2015}. In this limit,~\eqref{eq:S2} becomes
	\begin{align}
	S_2&=4\int\dif t\int\dif t' \chi_\textsc{a}(t)\chi_\textsc{b}(t')
	\Re{(\alpha_\textsc{a}^*\beta_\textsc{a}e^{\ii\Omega_\textsc{a}t})} \notag \\
	&\times
	\Re{\Big(\alpha_\textsc{b}^*\beta_\textsc{b}e^{\ii\Omega_\textsc{b}t'}\Big\langle [\phi(\bm{x}_\textsc{a},t),\phi(\bm{x}_\textsc{b},t')]\Big\rangle_{\!\!\rho_{_{0,\phi}}}\Big)}, \label{eq:S2_2}
	\end{align}
	
	For matter and cosmological constant-dominated universes in the case of minimal coupling of the field to the geometry, the field commutator between two events, $x=(\bm{x}_\textsc{a},t)$ and $x'=(\bm{x}_\textsc{b},t')$, is (see details in Appendix~\ref{appendix:commutators})
	\begin{multline}\label{eq:commutator}
	\langle[\phi(x),\phi(x')]\rangle_{\rho_\phi}=\frac{\ii}{4\pi}
	\Bigg[
	\frac{\delta(\Delta\eta+R)-\delta(\Delta\eta-R)}{a(t)a(t')R}\\
	+\frac{\theta(-\Delta\eta-R)-\theta(\Delta\eta-R)}{a(t)a(t')|\eta(t)\eta(t')|}
	\Bigg],
	\end{multline}
	where $\Delta\eta=\eta-\eta'=\eta(t)-\eta(t')$ and $R=\|\bm{x}_\textsc{a}-\bm{x}_\textsc{b}\|$. Immediately, due to the presence of the Heaviside $\theta$-function, we see that the support of the field commutator is not limited solely to boundaries of the light cone $\Delta\eta=\pm R$, and so we expect timelike signaling to be possible.
	
	Let us set the initial states of the detectors to be
	\begin{align}
	&\ket{\psi_{0,\textsc{a}}}=\frac{1}{\sqrt{2}}(\ket{e_\textsc{a}}-\ket{g_\textsc{a}}),\notag\\
	&\ket{\psi_{0,\textsc{b}}}=\frac{1}{\sqrt{2}}(\ket{e_\textsc{b}}+\ii\ket{g_\textsc{b}}).\label{eq:initstates}
	\end{align}
	We make this choice since it maximizes the signaling estimator~\eqref{eq:S2_2} in the case of zero gap detectors. Nevertheless, this choice is arbitrary, and any other initialization of detectors would lead to the same qualitative results.
	
	Using the initial detector states~\eqref{eq:initstates} and the field commutator~\eqref{eq:commutator}, the expression for $S_2$~\eqref{eq:S2} becomes
	\begin{equation}\label{eq:S2_final}
	S_2=\frac{1}{4\pi}(I_\delta+I_\theta),
	\end{equation}
	where
	\begin{align}
	I_\delta&=\frac{1}{R}\int_{\eta_{i\textsc{b}}}^{\eta_{f\textsc{b}}}\!\dif\eta\,\chi_A(\eta-R)
	\cos[\Omega_\textsc{b}t(\eta)] \cos[\Omega_\textsc{a}t(\eta-R)], \label{eq:Idelta}\\
	I_\theta&=\int_{\eta_{i\textsc{b}}}^{\eta_{f\textsc{b}}}\frac{\dif\eta_2}{|\eta_2|}
	\theta[\min(\eta_{f\textsc{a}},\eta_2-R)-\eta_{i\textsc{A}}]\cos[\Omega_\textsc{b}t(\eta_2)] \notag\\
	&\times\int_{\eta_{i\textsc{A}}}^{\min(\eta_{f\textsc{a}},\eta_2-R)}\frac{\dif\eta_1}{|\eta_1|}
	\cos[\Omega_\textsc{a}t(\eta_1)]. \label{eq:Itheta}
	\end{align}
	Here $\eta_{i\nu}=\eta(T_{i\nu})$ and $\eta_{f\nu}=\eta(T_{f\nu})$.
	
	Let us analyze the simple communication protocol laid out in~\cite{Blasco2015,Jonsson2015}: Alice encodes the bit ``1" by coupling her detector to the field at time $T_{i\textsc{a}}$ and decoupling at time $T_{f\textsc{a}}=T_{i\textsc{a}}+\Delta$, and the bit ``0" by remaining uncoupled. To later decode the message, Bob couples to the field at time $T_{i\textsc{b}}$, decouples at time $T_{f\textsc{b}}=T_{i\textsc{b}}+\Delta$, and measures his energy eigenstate. If he is excited, he interprets it as ``1", and ``0" otherwise. Notice that for simplicity we are keeping Alice and Bob's detectors switched on for an equal proper time interval $\Delta$, and recall that we are considering sudden switching of detectors, as given in~\eqref{eq:switching}. The number of bits per use of this binary communication channel that Alice can transmit to Bob is given by the Shannon capacity~\cite{Silverman1955}, which was shown in~\cite{Jonsson2015} to be
	\begin{equation}\label{eq:cc_general}
	C=\frac{\lambda_\textsc{a}^2\lambda_\textsc{b}^2}{8\ln 2}
	\left(\frac{S_2}{|\alpha_\textsc{b}||\beta_\textsc{b}|}\right)^2+\mathcal{O}(\lambda_\nu^6).
	\end{equation}
	For a matter or cosmological constant-dominated universe, with minimal coupling of the field to the curvature, and with initial detector states~\eqref{eq:initstates}, $S_2$ is given by~\eqref{eq:S2_final}. Hence the channel capacity~\eqref{eq:cc_general} becomes
	\begin{equation}\label{eq:cc}
	C=\frac{\lambda_\textsc{a}^2\lambda_\textsc{b}^2}{32\pi^2\ln2}
	(I_\delta+I_\theta)^2+\mathcal{O}(\lambda_\nu^6).
	\end{equation}
	We will study the form of the channel capacity~\eqref{eq:cc} when Alice and Bob are strictly timelike separated. The matter-dominated case was thoroughly analyzed in~\cite{Blasco2015,Blasco2016}. However, the cosmological constant-dominated scenario remains unexplored. Despite the mathematical similarities between them, we will show that there are fundamental and unintuitive physical differences in the abilities of timelike separated observers to communicate within the two cosmologies. Namely, timelike signals can carry considerably more information about the early universe when the spatial expansion is exponential as opposed to polynomial.
	
	\section{Timelike communication in polynomially and exponentially expanding cosmologies\label{sec:timelike_communication}}
	
	\begin{figure}[t]
		\includegraphics[width=0.50\textwidth]{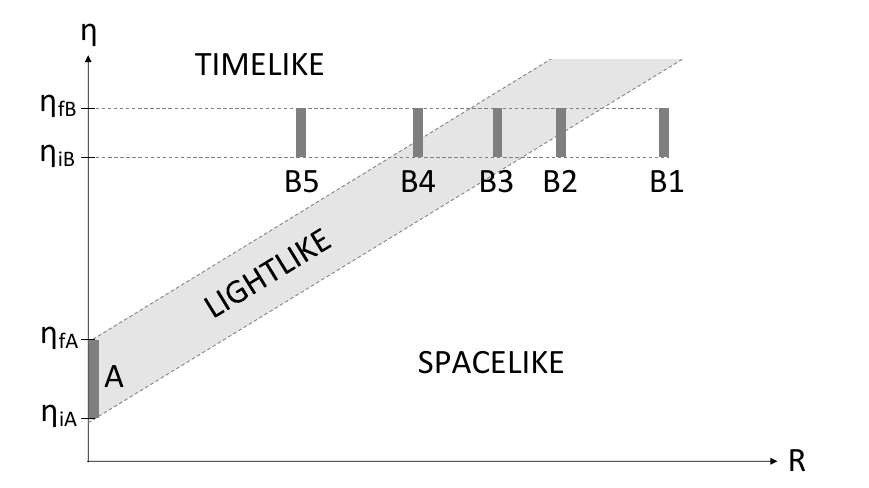}
		\caption{\label{fig:causal}The five possible causal relationships between the switching periods of Alice and Bob's detectors. In conformal time $\eta$ and comoving distance $R$, the boundaries of the light cones (diagonal lines) have slopes of $c=1$.}
	\end{figure}
	
	The form of Alice and Bob's communication channel capacity~\eqref{eq:cc} depends on the causal relationship between the supports of their switching functions. Fig.~\ref{fig:causal} shows the five possible causal relationships. When Bob is strictly spacelike separated from Alice, as in B1, both $I_\delta$~\eqref{eq:Idelta} and $I_\theta$~\eqref{eq:Itheta} in the channel capacity vanish, hence superluminal communication between the observers is indeed impossible. In the cases B2, B3 and B4, there is partial lightlike contact with Alice, so $I_\delta$ does not vanish entirely. As expected, lightlike communication is possible through a massless scalar field. Note that in the case of B4, the $I_\theta$ term also contributes to the channel capacity, meaning that communication is due to both lightlike and timelike signals. Timelike signaling is most evident when we consider detector B5, which is strictly within Alice's future light cone. Here, while $I_\delta$ vanishes, $I_\theta$ does not: in matter and cosmological constant-dominated universes with minimal coupling of the field to the curvature, slower than light communication is possible.
	
	\subsection{\label{sec:signal_timing}Signal timing}
	
	Before we study the channel capacity of timelike separated observers, let us discuss when Alice and Bob are timelike separated. We hold constant the switching times of Alice's detector, and ask the following question: where and when can Bob switch his detector on such that during his interaction with the field he is strictly within Alice's future light cone? That is, we fix $T_{i\textsc{a}}$ and $T_{f\textsc{a}}$ and look for the values of $T_{i\textsc{b}}$ and $R$ for which the two detectors, while switched on, are strictly timelike separated. From Fig.~\ref{fig:causal} it is evident that this occurs when 
	\begin{equation}
	\eta_{i\textsc{b}}> \eta_{f\textsc{a}}+R.
	\end{equation}
	If we keep the comoving separation between the detectors ($R$) constant, then
	\begin{equation}\label{eq:tib0}
	T_\text{min}^R=t(\eta(T_{i\textsc{a}}+\Delta)+R)
	\end{equation}
	is the smallest value of $T_{i\textsc{b}}$ at which there is strict timelike contact between the detectors. If instead we keep constant the time $T_{i\textsc{b}}$ at which Bob switches on his detector, then
	\begin{equation}\label{eq:R0}
	R_\text{max}=\eta(T_{i\textsc{b}})-\eta(T_{i\textsc{a}}+\Delta)
	\end{equation}
	is the largest comoving separation between Alice and Bob for which the two are fully timelike separated. One can trivially particularize $T_\text{min}^R$ and $R_\text{max}$ for the cosmologies generated by matter and a cosmological constant by using the appropriate forms of $\eta(t)$ from~\eqref{eq:w=0} and~\eqref{eq:w=-1}---and their inverses $t(\eta)$---in equations~\eqref{eq:tib0} and~\eqref{eq:R0}.
	
	The comoving distance, $R(t)$, is not usually the measure considered when discussing the spatial separation between us and distant cosmic objects. In astronomical terms, such separations are typically given in terms of the proper distance (i.e. the physical length of a measuring tape extended between us and the distant object as measured by us at time $t$). As a function of the comoving distance, the proper distance is given by
	\begin{equation}\label{eq:proper}
	P(t)=a(t)R(t).
	\end{equation}
	While the comoving distance between observers moving with the Hubble flow is independent of time, the proper distance between these observers increases as the universe expands. 
	
	Alternatively to what was done in~\cite{Blasco2015,Blasco2016}, instead of keeping the comoving distance between Alice and Bob constant, we can keep constant the proper distance. This requires at least one of Alice or Bob to be non-comoving. However, it is convenient to assume that both observers are comoving during their interaction time with the field, in order to obtain an analytic expression for the field commutator~\eqref{eq:commutator}. For this reason we approximate the channel capacity at a constant proper separation, $P$, by the capacity at a constant \textit{comoving} separation, \begin{equation}\label{eq:R0proper}
	R(T_{i\textsc{b}})=\frac{P}{a(T_{i\textsc{b}})}.
	\end{equation}
	This is a valid approximation as long as Alice and Bob's detector-field interaction times are much shorter than their temporal separation ($\Delta\ll T_{i\textsc{b}}-T_{f\textsc{a}}$). That is, we consider the expansion of the universe during the interaction time of the detectors with the field to be negligible, but we consider the full dynamics of the background spacetime between the emission and reception events. This is reasonable to expect if Bob is us and Alice is an early universe observer.
	
	The earliest time $T_\text{min}^P$ that Bob can switch on his detector while remaining strictly in Alice's timelike future and maintaining a constant proper separation $P$, is found by solving
	\begin{equation}\label{eq:tib0proper}
	\eta(T_\text{min}^P)=\eta(T_{i\textsc{a}}+\Delta)+\frac{P}{a(T_\text{min}^P)}.
	\end{equation}
	In a matter-dominated universe~\eqref{eq:w=0}, the solution is given by the single real root of the cubic equation
	\begin{equation}
	\left(T_\text{min}^P-\frac{R_0}{3}\right)^3=T_{f\textsc{a}}(T_\text{min}^P)^2,
	\end{equation}
	while in a cosmological constant-dominated universe~\eqref{eq:w=-1}, the solution to~\eqref{eq:tib0proper} is
	\begin{equation}
	T_\text{min}^R=\frac{1}{\lam}[\ln(1+P\lam)+T_{f\textsc{a}}].
	\end{equation}
	
	Finally, we can keep $T_{i\textsc{b}}$ constant and vary the proper distance between Alice and Bob. The largest value of $P$ for which the observers are strictly timelike separated is given by multiplying the comoving distance $R_\text{max}$~\eqref{eq:R0} by the appropriate scale factor~\eqref{eq:w=0} or~\eqref{eq:w=-1}.
	
	\subsection{\label{sec:cc}Channel capacity}
	
	The capacity of Alice and Bob's communication channel is given in expression~\eqref{eq:cc}. In the region of strict timelike contact of the detectors, the $I_\delta$ integral~\eqref{eq:Idelta} vanishes identically, while $I_\theta$~\eqref{eq:Itheta} becomes
	\begin{equation}\label{eq:Itheta2}
	I_{\theta}=
	\int_{\eta_{i\textsc{A}}}^{\eta_{f\textsc{a}}}\mspace{-15mu}\dif\eta_1\frac{\cos[\Omega_\textsc{a}t(\eta_1)]}{|\eta_1|}
	\int_{\eta_{i\textsc{b}}}^{\eta_{f\textsc{b}}}\mspace{-15mu}\dif\eta_2\frac{\cos[\Omega_\textsc{b}t(\eta_2)]}{|\eta_2|}.
	\end{equation}
	Changing the integration variable to comoving time one obtains
	\begin{equation}\label{eq:Itheta3}
	I_{\theta}=
	\int_{T_{i\textsc{a}}}^{T_{i\textsc{a}}+\Delta}\mspace{-20mu}\dif t_1 \frac{\cos(\Omega_\textsc{a}t_1)}{a(t_1)|\eta(t_1)|}
	\int_{T_{i\textsc{b}}}^{T_{i\textsc{b}}+\Delta}\mspace{-20mu}\dif t_2 \frac{\cos(\Omega_\textsc{b}t_2)}{a(t_2)|\eta(t_2)|}.
	\end{equation}
	We will particularize this expression to the two cosmologies that we are considering.	
	
	\subsubsection{\label{sec:cc_matter}Matter-dominated cosmology}
	
	In the matter-dominated universe ($w=0$), using~\eqref{eq:w=0} we obtain
	\begin{equation}\label{eq:a_eta_w=0}
	a(t)|\eta(t)|=3t,
	\end{equation}
	which is the proper \textit{particle horizon} of the observer at time $t$, i.e. the maximal proper distance that light could have traveled to the observer in the age of the Universe. Notice that in this case~\eqref{eq:a_eta_w=0} also corresponds to twice the Hubble radius at time $t$. For the case of non-zero gap detectors, $\Omega_\nu>0$, equation~\eqref{eq:Itheta3} becomes
	\begin{align}\label{eq:Itheta_w=0}
	I_{\theta}^{w=0}=\frac{1}{9}
	&(\Ci[\Omega_\textsc{a}(T_{i\textsc{a}}+\Delta)]-\Ci[\Omega_\textsc{a}T_{i\textsc{a}}]) \notag \\
	\times&(\Ci[\Omega_\textsc{b}(T_{i\textsc{b}}+\Delta)]-\Ci[\Omega_\textsc{b}T_{i\textsc{b}}]),
	\end{align}
	where $\Ci$ is the cosine integral function,
	\begin{equation}
	\Ci(z)=\int_z^\infty \dif t \frac{\cos t}{t}.
	\end{equation}
	If we assume that $\Delta\ll T_{i\textsc{a}}<T_{i\textsc{b}}$,~\eqref{eq:Itheta_w=0} simplifies to
	\begin{equation}\label{eq:Itheta_w=0_2}
	I_{\theta}^{w=0}\simeq \frac{\Delta^2}{9}\frac{\cos(\Omega_\textsc{a}T_{i\textsc{a}})}{T_{i\textsc{a}}}
	\frac{\cos(\Omega_\textsc{b}T_{i\textsc{b}})}{T_{i\textsc{b}}}.
	\end{equation}
	This assumption is reasonable since, as mentioned above, we expect the time scale of the detectors being switched on to be much smaller than the cosmological time scale on which the universe evolves.
	
	Note that taking the limit $\Omega_\nu\to 0$ in~\eqref{eq:Itheta_w=0} one obtains
	\begin{equation}\label{eq:Itheta_gapless}
	\lim_{\Omega_\nu\to 0}
	\left(I_{\theta}^{w=0}\right)=
	\frac{1}{9}
	\ln\left(\frac{T_{i\textsc{a}}+\Delta}{T_{i\textsc{a}}}\right)
	\ln\left(\frac{T_{i\textsc{b}}+\Delta}{T_{i\textsc{b}}}\right),
	\end{equation}
	which is the result derived in~\cite{Blasco2016} for gapless detectors. 
	
	Therefore, for strictly timelike separated observers in the matter-dominated universe, the channel capacity~\eqref{eq:cc} becomes
	\begin{align} \label{eq:cc_w=0}
	C^{w=0}_{\Omega_\nu>0}&=\frac{\lambda_\textsc{a}^2\lambda_\textsc{b}^2}{2592\pi^2\ln2}
	\left(\Ci[\Omega_\textsc{a}(T_{i\textsc{a}}+\Delta)]-\Ci[\Omega_\textsc{a}T_{i\textsc{a}}]\right)^2 \notag \\
	&\mspace{96mu}
	\times \left(\Ci[\Omega_\textsc{b}(T_{i\textsc{b}}+\Delta)]-\Ci[\Omega_\textsc{b}T_{i\textsc{b}}]\right)^2 \notag \\
	&\simeq\frac{\lambda_\textsc{a}^2\lambda_\textsc{b}^2\Delta^4}{2592\pi^2\ln2}
	\left(\frac{\cos(\Omega_\textsc{a}T_{i\textsc{a}})}{T_{i\textsc{a}}}
	\frac{\cos(\Omega_\textsc{b}T_{i\textsc{b}})}{T_{i\textsc{b}}}\right)^2, \\
	C^{w=0}_{\Omega_\nu=0}&=\frac{\lambda_\textsc{a}^2\lambda_\textsc{b}^2}{2592\pi^2\ln2}
	\left(
	\ln\left(\frac{T_{i\textsc{a}}+\Delta}{T_{i\textsc{a}}}\right)
	\ln\left(\frac{T_{i\textsc{b}}+\Delta}{T_{i\textsc{b}}}\right)
	\right)^2\!\!\!, \notag
	\end{align}
	where we used equations~\eqref{eq:Itheta_w=0},~\eqref{eq:Itheta_w=0_2} and~\eqref{eq:Itheta_gapless} for $I_\theta^{w=0}$.
	
	\subsubsection{\label{sec:cc_lambda}\texorpdfstring{$\Lambda$}{}-dominated cosmology}
	
	In the cosmological constant-dominated universe (\mbox{$w=-1$}), we see from~\eqref{eq:w=-1} that the denominators of the integrands of~\eqref{eq:Itheta3} become
	\begin{equation}\label{eq:a_eta_w=-1}
	a(t)|\eta(t)|=\frac{1}{\lam},
	\end{equation}
	which in this case coincides with both the Hubble radius and the proper \textit{event horizon} of the observer at time $t$, i.e. the proper distance that light emitted at time $t$ would travel in the lifetime of the Universe. Critically, as opposed to the particle horizon in the matter-dominated universe~\eqref{eq:a_eta_w=0}, equation~\eqref{eq:a_eta_w=-1} does not depend on time. In the $\Lambda$-dominated cosmology,~\eqref{eq:Itheta3} becomes
	\begin{align}
	I_{\theta}^{w=-1}&=\frac{4|\Lambda|}{\Omega_\textsc{a}\Omega_\textsc{b}}
	\sin\left(\frac{\Omega_\textsc{a}\Delta}{2}\right)
	\sin\left(\frac{\Omega_\textsc{b}\Delta}{2}\right) \notag \\
	&\times\cos\left[\Omega_\textsc{a}\left(T_{i\textsc{a}}+\frac{\Delta}{2}\right)\right]
	\cos\left[\Omega_\textsc{b}\left(T_{i\textsc{b}}+\frac{\Delta}{2}\right)\right],
	\label{eq:Itheta_w=-1}
	\end{align}
	and the channel capacity~\eqref{eq:cc} acquires the form,
	\begin{align}
	C^{w=-1}&=
	\frac{\lambda_\textsc{a}^2\lambda_\textsc{b}^2\Lambda^2}{162\pi^2\ln2\,\Omega_\textsc{a}^2\Omega_\textsc{b}^2}
	\sin^2\left(\frac{\Omega_\textsc{a}\Delta}{2}\right)
	\sin^2\left(\frac{\Omega_\textsc{b}\Delta}{2}\right) \notag \\
	&\times \cos^2\left[\Omega_\textsc{a}\left(T_{i\textsc{a}}+\frac{\Delta}{2}\right)\right]
	\cos^2\left[\Omega_\textsc{b}\left(T_{i\textsc{b}}+\frac{\Delta}{2}\right)\right]
	\label{eq:cc_w=-1}.
	\end{align}
	We are now ready to compare the abilities of timelike separated observers to communicate within the two cosmologies.
	
	\subsection{\label{sec:results}Results}
	
	Let us now compare the communication channel capacities between an early-universe signal emitter, Alice, and a late-time receiver, Bob, in cosmologies generated by matter and a cosmological constant.
	
	We will focus on the channel capacities when Alice and Bob's detectors are strictly timelike separated. We recall that, since the real quanta of the massless scalar field travel at the speed of light, one may intuitively have expected that the channel capacities in this causal regime are zero. However, as explained above, this is the relevant case of \textit{quantum collect calling}, where slower than light communication through the massless scalar field is possible if Alice and Bob's detectors are prepared in coherent superpositions of their excited and ground states in scenarios where the strong Huygens principle is violated~\cite{Jonsson2015,Blasco2016}, as is the case for minimally coupled fields in FRW backgrounds  \cite{Blasco2015,Blasco2016}.
	
	The initial states of the qubit detectors with which Alice and Bob couple to the field are defined in~\eqref{eq:initstates}. To facilitate a comparison with the results in~\cite{Blasco2016,Blasco2015}, the initialization is chosen to maximize the channel capacity in the case of zero-gap detectors. We will not particularize to the zero-gap case, but we will for simplicity consider the energy gaps of the two detectors to be equal: $\Omega_\textsc{a}=\Omega_\textsc{b}=\Omega$. Recall that the detectors are switched on and off suddenly, according to~\eqref{eq:switching}.
	
\begin{figure}[t]
		\includegraphics[width=0.45\textwidth]{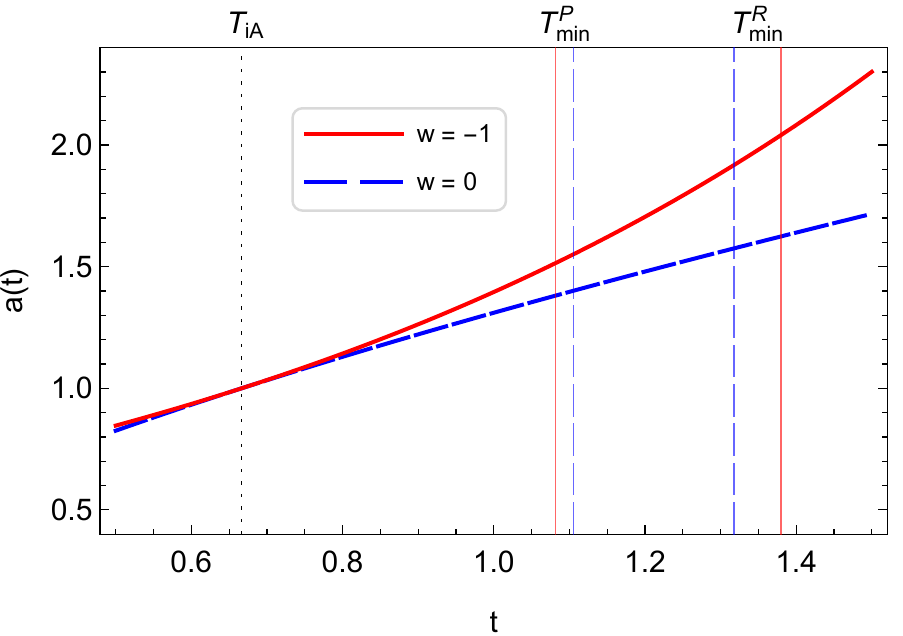}
		\caption{\label{fig:bounds}Scale factors governing the expansion of the matter ($w=0$) and cosmological constant ($w=-1$) dominated universes, plotted as functions of time. The four rightmost vertical lines show the earliest times $T_{i\textsc{b}}$ for which Alice and Bob are strictly timelike separated, while keeping constant their spatial separation. When comoving separation is held constant, $R=1/2$, and when proper separation is approximated as constant, $P(T_{i\textsc{b}})=1/2$. Here $T_{i\textsc{a}}=2/3$, $\Delta=1/100$, $\kappa_1=1/4$, $\kappa_2=\exp(-2/3)$ and $\lam=1$.}
	\end{figure}

	To elucidate the effects of cosmological expansion on the ability of observers to communicate, we use our freedom to choose a reference scale for the constant factors to set the two spacetimes and their rates of expansion to be equal at a given initial time, which in our case will be the time at which Alice's detector is switched on, $T_{i\textsc{a}}$. To that effect, we set:
	\begin{align}
	&a_{w=0}(T_{i\textsc{a}})=a_{w=-1}(T_{i\textsc{a}})=1, \notag \\
	&\dot{a}_{w=0}(T_{i\textsc{a}})=\dot{a}_{w=-1}(T_{i\textsc{a}})=1. \label{eq:init_cond_a}
	\end{align}
	This is done by setting $\kappa_1=1/4$, $\kappa_2=\exp(-2/3)$, $\lam=1$ and $T_{i\textsc{a}}=2/3$. The effects of this choice in both dynamics can be seen in Fig.~\ref{fig:bounds}.
	
	The comoving time for which each detector is switched on is set to $\Delta=1/100$. For the values of $T_{i\textsc{b}}$ that we will work with, this ensures that \mbox{$\Delta\ll T_{i\textsc{b}}-T_{f\textsc{a}}$}, which we require in order to approximate a constant proper separation between the detectors, as discussed in Sec.~\ref{sec:signal_timing}. 
	
	\begin{figure}[t]
		\centering
		\subfigure{
			\includegraphics[width=0.45\textwidth]{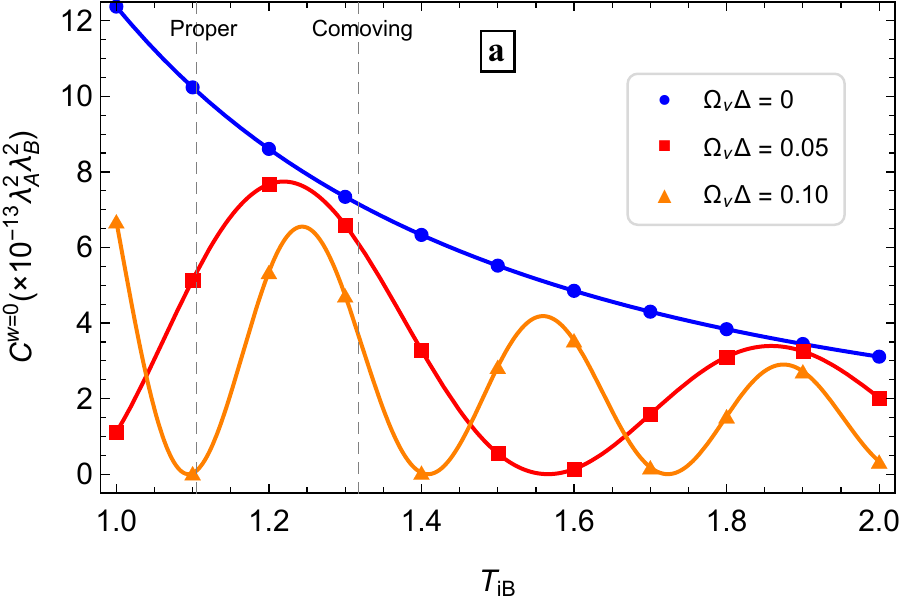}}
		\subfigure{
			\includegraphics[width=0.45\textwidth]{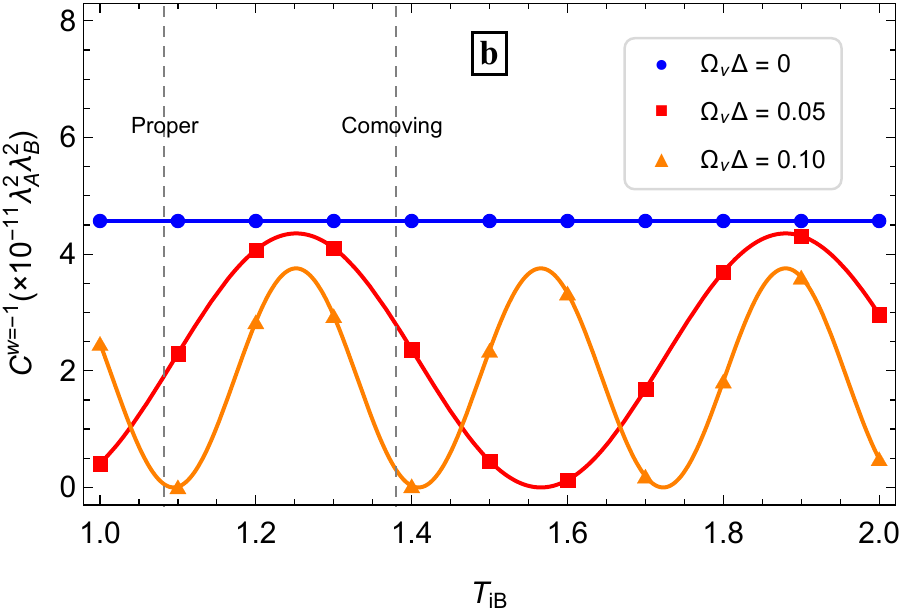}}
		\caption{ Variations of the channel capacity with the instant $T_{i\textsc{b}}$ when Bob's detector is switched on in a universe dominated by a) matter and b) a cosmological constant. We study the case when the detectors are strictly timelike separated: the vertical lines indicate the earliest $T_{i\textsc{b}}$ for which this occurs while the proper/comoving separation is kept constant. Notice that the plots only show the correct channel capacity for values of $T_{i\textsc{b}}$ that ensure timelike separation (to the right of each dashed vertical line) since \eqref{eq:Itheta_w=0} and \eqref{eq:Itheta_w=-1} are only valid for timelike separation between Alice and Bob. Here $T_{i\textsc{a}}=2/3$, $\Delta=1/100$, $\kappa_1=1/4$, $\kappa_2=\exp(-2/3)$ and $\lam=1$. When comoving separation is held constant, $R=1/2$. When proper separation is approximated as constant, $P(T_{i\textsc{b}})=1/2$. Various values of the detectors' energy gap $\Omega$ are considered.}
		\label{fig:cc}
	\end{figure}
	
	We see from equations~\eqref{eq:cc_w=0} and~\eqref{eq:cc_w=-1} that the timelike channel capacities in the two cosmologies are both independent of the distance (proper or comoving) separating Alice and Bob. It was pointed out in~\cite{Blasco2015,Blasco2016} (for the matter-dominated case) that this fact allows timelike channels to potentially convey more information from spatially distant events than light signals due to the fact that the timelike channel capacity does not decay with the distance to the source. We see that this is also true in the cosmological constant-dominated universe. 
	
    Remarkably, we find critical differences in the ability of Alice to communicate with Bob through timelike channels in the two cosmologies. Namely, the timelike \textit{afterglow} of Alice's interaction with the field remains constant (up to oscillations) before reaching Bob in the $\Lambda$-dominated universe, in stark contrast with the time decay present in the matter-dominated case.
	
	This can be seen in Fig.~\ref{fig:cc}, where we plot the channel capacities as functions of the instant that Bob's detector is switched on, $T_{i\textsc{b}}$. We consider two different situations: 1) constant comoving separation between Alice and Bob, $R=1/2$, as in~\cite{Blasco2015,Blasco2016}, and 2) constant proper separation, $P(T_{i\textsc{b}})=1/2$. We see that there are no relevant qualitative differences between the situations 1) and 2). The only difference in communication that the choice of  distance measure effects, is the values of $T_{i\textsc{b}}$ for which the detectors are strictly timelike separated. Namely, keeping constant the proper separation results in strictly timelike separated detectors at lower $T_{i\textsc{b}}$ than when maintaining the same comoving distance constant. This is due to our choice of reference scale when normalizing the scale factors~\eqref{eq:init_cond_a}: since $a(T_{i\textsc{b}})>1$, at time $T_{i\textsc{b}}$ a given comoving separation is physically larger (and hence takes light longer to traverse) than the same proper separation. The distance measure that we choose to keep constant therefore affects the relative spacetime positioning of Alice and Bob, displacing the positions of the timelike connected regions in Fig.~\ref{fig:cc}.
	
	Notice that the magnitudes of the channel capacities in Fig.~\ref{fig:cc}a are much smaller than those reported in~\cite{Blasco2016}. This is mainly due to us considering a detector-field interaction time, $\Delta$, that is several orders of magnitude less than that in~\cite{Blasco2016} (note from~\eqref{eq:cc_w=0} that $C^{w=0}\propto\Delta^4$). Indeed, as expected, the longer Alice interacts with the field, the more information she encodes for Bob to later recover.
	
	If we look at Fig.~\ref{fig:cc}a, we see that in the matter-dominated universe the channel capacity has a polynomial decay in time: Bob's ability to retrieve Alice's signal is suppressed the longer he waits to do so.
	
	Remarkably, Fig.~\ref{fig:cc}b shows that the channel capacity in the exponentially expanding cosmology does not decay as the time that Bob waits to read out the signal increases: even if Bob waits the age of the universe to recover the signal, the channel capacity between him and Alice will remain the same (up to oscillations).
	
	The behaviour shown in Fig.~\ref{fig:cc}, stems from the time dependence of the equations for the channel capacities in the matter and $\Lambda$-dominated cosmologies,~\eqref{eq:cc_w=0} and~\eqref{eq:cc_w=-1}, respectively. If, for illustration, we look at the approximated form of~\eqref{eq:cc_w=0} (which applies to the results in the figures since \mbox{$\Delta\ll T_{i\textsc{a}}<T_{i\textsc{b}}$}), we see that $C^{w=0}\propto T_{i\textsc{b}}^{-2}$. The capacity in the $\Lambda$-dominated case exhibits no such decay with $T_{i\textsc{b}}$. 
	
	This result seems contrary to the physical intuition that, since an exponential expansion is faster than a polynomial one, the information encoded in the field by Alice in the former case should get dispersed more, resulting in a faster decaying channel capacity, as is the case with lightlike signals. What is more, not only does the channel capacity in the $\Lambda$-dominated cosmology not decay, but it actually grows as $\Lambda^2$, meaning that more information can in principle be broadcast from Alice to Bob the faster the exponential expansion of the Universe is. 
	
	Along with the decay (or lack thereof) discussed above, both channel capacities also exhibit oscillations with $T_{i\textsc{a}}$ and $T_{i\textsc{b}}$ at frequencies equal to the energy gap of the detectors, $\Omega$.
	
	\section{Conclusions\label{sec:conclusions}}
	
	By using the protocols of \textit{quantum collect calling}~\cite{Blasco2016,Jonsson2015} it is possible to detect signals broadcast by early universe observers in our timelike past (when there is no light contact), greatly increasing both the volume of observable spacetime and the amount of recoverable information from that available through classical observational methods. This ability of timelike separated observers to communicate is fundamentally dependent on 1) the coupling of the field to the underlying geometry, 2) the dimensionality of spacetime, and 3) the geometry of spacetime. We focused here on the case of minimal coupling in (3+1)-dimensions, which was shown in~\cite{Blasco2016, Blasco2015} to be a viable setup for timelike signaling in the case of a polynomially-expanding, matter-dominated cosmology. In this paper, we have analyzed the exponentially-expanding, cosmological constant-dominated universe, and we found unexpected fundamental differences between the two cases.
	
	We quantified the ability of timelike separated observers, Alice and Bob, to exchange information in the two cosmologies. To do so, we computed a lower bound to the Shannon capacity of the  channel established when they communicate using antennas coupled to the quantum field. We showed that, as in the matter-dominated cosmology, the channel capacity in the $\Lambda$-dominated case is independent of the spatial and temporal separations between Alice and Bob. 
	
Most interestingly, we also found that in the exponentially expanding universe, there is no decay of the channel capacity with Alice and Bob's individual coupling times. This means that Bob can wait as long as he wants and the amount of information that he can recover from Alice will not change. What is more, we find that the channel capacity is proportional to $\Lambda^2$. This implies that the faster the expansion of the Universe is, the greater the ability of Bob to recover the information sent by Alice through timelike communication.
    
This is contrary to the polynomial decay present in the matter-dominated universe and studied in previous literature~\cite{Blasco2015,Blasco2016}, and it challenges the---perhaps intuitive---physical expectation that a faster spatial expansion results in less information reaching an observer, since it would be dispersed as the Universe expands.
    
The unintuitive lack of decay in a $\Lambda$-dominated cosmology is made even more interesting when we note that our own Universe seems to have been exponentially expanding at very early times in its history, and appears to be currently dominated by a cosmological constant as well. This opens up fascinating possibilities of applying the theory presented here, at least in principle, to observe our distant timelike past, or to send signals to observers in our timelike future.

	\section*{Acknowledgments}
	
    The authors are very thankful to Ana Blasco, Luis J. Garay and Mercedes Mart\'in-Benito for helpful discussions and invaluable feedback. The authors gratefully acknowledge the financial support provided by the NSERC Discovery and USRA Programs.

	\appendix
	
	\section{Field commutator for minimal coupling of the field to the geometry}\label{appendix:commutators}
	
	In this appendix, we review the calculations originally outlined in~\cite{Blasco2015,Blasco2016,Poisson2011}.
	We start with the expression for the expectation value of the field commutator between two events, $x=(\bm{x}_\textsc{a},t)$ and $x'=(\bm{x}_\textsc{b},t')$, in terms of the advanced and retarded Green functions, $G_-$ and $G_+$, respectively:
	\begin{equation}\label{eq:commutator_green}
	   \langle[\phi(x),\phi(x')]\rangle=\ii\frac{G_-(x,x')-G_+(x,x')}{4\pi}.
	\end{equation}
	The $G_\pm$ are solutions to the wave equation~\eqref{eq:wave_eqn} with a point-like source
	\begin{equation}\label{eq:wave_eqn2}
	    (\Box-\xi\mathcal{R})G_\pm(x,x')=-\frac{4\pi}{a(\eta)^4}\delta(\eta-\eta')\delta^3(\bm{x}-\bm{x}').
	\end{equation}
	Rescaling by $a(\eta)a(\eta')$ and introducing the Fourier transform $\hat{g}$, we can rewrite $G_\pm$ as
	\begin{equation}
	    G_\pm(x,x')=\frac{\pm\theta(\pm\eta\mp\eta')}{(2\pi)^3 a(\eta)a(\eta')}\int\d[3]{\bm{k}}e^{\ii\bm{k}\cdot(\bm{x}-\bm{x}')}\hat{g}(\eta,\eta',k),
	\end{equation}
	which upon substitution into~\eqref{eq:wave_eqn2} gives the auxiliary differential equation
	\begin{equation}\label{eq:wave_eqn3}
	    \left(\frac{\d[2]}{\dif\eta^2}+k^2-(1-6\xi)\frac{\alpha^2-1/4}{\eta^2}\right)\hat{g}(\eta,\eta',k)=0,
	\end{equation}
	with boundary conditions
	\begin{equation}
	    \hat{g}(\eta=\eta',k)=0,\quad \frac{\dif\hat{g}}{\dif\eta}(\eta=\eta',k)=4\pi.
	\end{equation}
	Here, we have defined $\alpha=\left|(3-3w)/(6w+2)\right|$, where we recall that $w=p/\rho$ is the pressure to density ratio of the perfect fluid generating our spacetime.
	
	In the case of minimal coupling, $\xi=0$. Then, the solution $\hat{g}_\alpha(\eta,\eta',k)$ (where we have explicitly denoted the $\alpha$ dependence) to~\eqref{eq:wave_eqn3} is given by equation (55) in~\cite{Blasco2016}. The commutator~\eqref{eq:commutator_green} then becomes
	\begin{align}\label{eq:commutator_int}
	    \langle[\phi(x),\phi(x')]\rangle&=\ii\frac{\theta(-\Delta(\eta))-\theta(\Delta(\eta))}{\pi^2 a(t)a(t')R} \notag \\
	    &\times \int_0^\infty\dif k\sin(kR)\hat{g}_\alpha(\eta(t),\eta(t'),k),
	\end{align}
	where $\Delta(\eta)=\eta(t)-\eta(t')$ and $R=\|\bm{x}_\textsc{a}-\bm{x}_\textsc{b}\|$. In both matter ($w=0$) and $\Lambda$ ($w=-1$) dominated cosmologies, $\alpha=3/2$, and the integral in~\eqref{eq:commutator_int} can be solved analytically, yielding expression~\eqref{eq:commutator}.

	\bibliography{references}
	\bibliographystyle{apsrev4-1}

\end{document}